\documentclass[12pt,preprint]{aastex}
\begin{document}

\parindent=1.0cm

\title 
{The Stellar Content of the Post-Starburst S0 Galaxy NGC 5102
\altaffilmark{1} \altaffilmark{2}}

\author{T. J. Davidge}

\affil{Herzberg Institute of Astrophysics,
\\National Research Council of Canada, 5071 West Saanich Road,
\\Victoria, B.C. Canada V9E 2E7\\ {\it email: tim.davidge@nrc.ca}}

\altaffiltext{1}{Based on observations obtained at the
Gemini Observatory, which is operated by the Association of Universities
for Research in Astronomy, Inc., under a co-operative agreement with the
NSF on behalf of the Gemini partnership: the National Science Foundation
(United States), the Particle Physics and Astronomy Research Council
(United Kingdom), the National Research Council of Canada (Canada),
CONICYT (Chile), the Australian Research Council (Australia), CNPq (Brazil),
and CONICET (Argentina).}

\altaffiltext{2}{This publication makes use of data products
from the Two Micron All Sky Survey, which is a joint project of the University of
Massachusetts and the Infrared Processing and Analysis Center/California
Institute of Technology, funded by the National Aeronautics and Space Administration
and the National Science Foundation.}

\begin{abstract}

	The stellar content of the S0 galaxy NGC 5102 
is investigated using deep $r'$ and $i'$ images obtained with GMOS-S. 
A modest population of bright main sequence stars and red 
supergiants (RSGs) is detected throughout the western portion of the disk. 
Based on the numbers of main sequence stars, the 
star formation rate (SFR) in NGC 5102 during the past ten million years 
is estimated to have been 0.02 M$_{\odot}$ year$^{-1}$. 
The majority of red giant branch (RGB) stars in the disk of 
NGC 5102 have [M/H] between --0.9 and --0.1, and the metallicity distribution of 
RGB stars at intermediate galactocentric radii peaks near [M/H] $\sim -0.6$. 
RGB stars are traced out to galactocentric distances of 10 kpc, which corresponds to 
$\sim 14$ disk scale lengths. A large population of bright asymptotic giant branch (AGB) 
stars are seen throughout the western portion of the disk, and the youngest of these have 
log(t$_{yr}) \sim 8.1$. It is concluded that (1) stars that formed within the 
past Gyr comprise $\sim 20\%$ of the total stellar disk mass, and (2) the SFR during 
intermediate epochs in the disk of NGC 5102 was at least 1.4 M$_{\odot}$ year$^{-1}$. 
Thus, large-scale star formation occured throughout the disk of NGC 5102 at approximately 
the same time that similar elevated levels of star formation occured in the bulge. 
It is suggested that NGC 5102 was a spiral galaxy that experienced a galaxy-wide 
episode of star formation that terminated a few hundred Myr in the past, and that 
much of its interstellar medium was ejected in an outflow.

\end{abstract}


\section{INTRODUCTION}

	NGC 5102 is a member of the Centaurus Group of galaxies and exhibits 
the classical morphological characteristics of an S0 galaxy. Still, 
NGC 5102 has other properties that are not typical of the S0 class. For example, NGC 5102 
has bluer colors and the spectrum has deeper Balmer absorption lines (Gallagher, 
Faber, \& Balick 1975) than is typical for this type of object. 
NGC 5102 is also one of the most HI-rich S0s known. Much of the HI is 
in an annulus centered at 3.5 arcmin radius, and such an HI distribution 
is unique among S0s (van Woerden et al. 1993). While the dust 
temperature of NGC 5102 inferred from IRAS observations 
is not peculiar for S0s, the overall dust mass is low (van Woerden et al. 1993).
Kraft et al. (2005) find a low number of x-ray binaries 
in NGC 5102 when compared with other early-type galaxies, and further note that the 
specific frequency of globular clusters is lower than in other S0s. 

	A number of studies have examined the stellar content in the central 
regions of NGC 5102. The nucleus of NGC 5102 is redder than its immediate surroundings, 
and is an x-ray source, suggesting that it may harbour a low-luminosity AGN (Kraft 
et al. 2005). The central regions (van den Bergh 1976; Pritchet 1979) and 
bulge (Pritchet 1979) of NGC 5102 have blue colors, and there are spectroscopic 
signatures of A and B stars (Gallagher et al. 1975; Rocca-Volmerange \& Guiderdoni 1987). 
Deharveng et al. (1997) resolved UV-bright stars near the 
center of NGC 5102. These probably formed $\sim 15$ Myr in the past, although 
Deharveng et al. (1997) note that older post asymptotic giant branch (PAGB) stars 
also have photometric properties that are similar to those of the observed objects. 
van den Bergh (1976) discovered a ring of H$\alpha$ emission. The ring has a diameter 
of $\sim 1.3$ kpc (McMillan, Ciardullo, \& Jacoby (1994) which is similar in size to 
supershells in other galaxies. The center of the ring is consistent 
with an origin in the bulge, rather than the nucleus, and 
the expansion velocity is indicative of an age of only $\sim 10^7$ years 
(McMillan et al. 1994). The central regions of NGC 5102 were also an area of active 
star formation during intermediate epochs. Bica (1988) concludes from 
visible wavelength spectra that roughly 54\% of the central light in $V$ originates from 
a population with an age $\sim 500$ Myr, while 
Kraft et al. (2005) model the UV-to-red spectrum and conclude that at least 55\% of the 
stellar content originates from an intermediate-age population with an age $< 3$ Gyr. 

	Recent star formation has not been restricted to the spheroid of NGC 5102.
van den Bergh (1976) identified groups of young stars associated with 
HII regions in the south east portion of the NGC 5102 disk, while Danks, Laustsen, \& 
van Woerden (1979) found a number of stars that may be supergiants. 
McMillan et al. (1994) detected 7 HII regions in the disk, 
five of which are located to the south east of the galaxy center, 
while two are located close to the major axis in the south west quadrant. 

	The deepest photometric investigation of the NGC 5102 disk was conducted by 
Karachentsev et al. (2002), who used WFPC2 images to construct an $(I, V-I)$ CMD of stars 
in the north east portion of the galaxy. The WFPC2 CMD is dominated by 
a broad RGB that is indicative of a spread in age and/or metallicity. A spray of AGB stars 
is also seen. Prominent blue or red supergiant (RSG) sequences are not evident in 
these data, although the region sampled is far removed from the area that 
contains HII regions at the present day.

	Studies of the spatial distribution of stars that formed during the past $\sim 
1$ Gyr will provide insights into the recent evolution of NGC 5102, and may 
provide hints as to its evolutionary status in the context of other galaxies in the 
local Universe. There are basic issues that can be addressed with moderately wide 
field imaging studies. For example, is there evidence for recent star formation 
in parts of the disk other than those that currently contain bright HII regions? 
Moreover, did the extensive star-forming activity 
that occured during intermediate epochs in the bulge extend into the 
disk? If evidence for wide-spread elevated star formation rates (SFRs) during 
intermediate epochs are seen throughout the disk then this argues for a galaxy-wide star 
formation trigger. In contrast, if elevated SFRs were restricted to the innermost regions 
of the galaxy during the past few Gyr then this argues for a less dramatic triggering 
mechanism, such as the buckling of a bar or perhaps the assimilation of a 
small, gas-rich companion. 

	In the current paper, deep $r'$ and $i'$ images obtained with GMOS-S 
are used to investigate the stellar content of 
the western portion of the NGC 5102 disk. The $r'$ and $i'$ filters were 
selected for this program as they are well-suited to the detection of the bright 
red AGB stars that are conspicuous signatures of star formation during 
intermediate epochs, as well as the red RGB stars that belong to the moderately 
metal-rich populations that frequent the disks of spiral galaxies. 
While observations through the $g'$ filter may have given more metallicity 
sensitivity, the integration times needed to detect the reddest AGB and RGB stars 
in $g'$ are prohibitive.

	The paper is structured as follows. Details of the observations 
and the procedures used to reduce the data can be found in \S 2, while the
color-magnitude diagrams (CMDs) constructed from the 
data are presented in \S 3. The properties of the young, old, and intermediate age 
components are the subjects of \S\S 4 -- 6. A summary and discussion of the results 
follows in \S 7.

\section{OBSERVATIONS \& REDUCTION}

	NGC 5102 was observed with GMOS (Crampton et al. 2000) on the Gemini South 
telescope as part of queue program GS-2006A-Q-45. The GMOS detector is a mosaic of 
three $2048 \times 4068$ EEV CCDs. The raw sampling on the sky is $0.072$ arcsec 
pixel$^{-1}$, and the output was binned $2 \times 2$ pixels$^2$ during read-out.
This binning allowed the final spatial sampling to better match the delivered image 
quality, which was measured from stars in the processed images to be 0.6 ($i'$) 
and 0.7 ($r'$) arcsec FWHM. 

	A field that is centered at 13$^h$ 10$^m$ 40$^s$ Right Ascension and 
--36$^o$ 36$^{'}$ 51$^{''}$ Declination (E2000), and therefore samples the 
western half of the NGC 5102 disk, was observed through Sloan Digital Sky Survey (SDSS) 
$r'$ and $i'$ filters (Fukugita et al. 1996). Transmission curves 
for the filters used in GMOS-S are available at the Gemini website 
\footnote[3]{http://www.gemini.edu/sciops/instruments/gmos/filters/GMOSS\_filters.gif}. 
The area that was observed is marked in Figure 1, which shows a 
section of the Digitized Sky Survey centered on NGC 5102. Ten 450 sec 
exposures were recorded filter$^{-1}$, and so the total exposure 
time is 4500 sec filter$^{-1}$.

	A standard pipeline for the processing of visible wavelength imaging observations 
was applied to the data. The output from each CCD was scaled to a common gain, 
and the results were mosaiced to form a single image for each exposure. These were then
bias-subtracted and flat-fielded. Prominent interference 
fringes are seen in the flat-fielded $i'$ images, 
and a fringe frame was constructed by median-combining flat-fielded $i'$ images of the 
various Centaurus Group galaxies that were observed as part of this program. The target 
galaxies were intentionally positioned on different parts of the science detector to 
facilitate the suppression of galaxy light when the images of 
all the targets are combined to make the fringe frame. The resulting 
fringe frame was then subtracted from the flat-fielded NGC 5102 $i'$ images. 
The telescope pointing was offset in a five point dither pattern between exposures to 
facilitate the filling of the gaps between the CCDs in the final processed images. 
The flat-fielded $r'$ and fringe subtracted $i'$ images were 
spatially registered to correct for the dither offsets, and the median intensity 
at each aligned pixel location was computed for each filter. The final step was to 
trim the stacked images to the area on the sky that is common to all exposures.

	Landolt (1992) standard stars are observed with GMOS throughout each semester 
to provide calibration information for queue-scheduled GMOS imaging observations. For the 
current study, standard star images that were recorded within a few weeks of the NGC 5102 
observations were fetched from the Gemini data archive and 
these were reduced using the procedures described above. 
While standard stars were not observed for this 
program on the same nights as the observations, the sky transparency was 
verified by the observering team at the telescope by checking 
satellite weather information, the brightness of the guide star, and 
the exposure-to-exposure photometric stability of stars in the target field.
It is also standard practice for the observing team to check visually sky 
conditions throughout the night.

	The standard star magnitudes listed by Landolt (1992) were 
transformed into the SDSS system using the relations defined by Smith et al. (2002), 
so that the photometric calibration is in the SDSS system. The instrumental brightnesses 
of the standard stars were measured from the reduced images in apertures that 
sampled the stellar PSF past the point where it blended with the sky background. 
The standard star observations were made near the meridian, 
and so extinction coefficients could not be calculated. Lacking a fiducial extinction 
curve for the Gemini South site, it was decided to adopt 
k$_{r'} = 0.10$ and k$_{i'} = 0.07$ mag airmass$^{-1}$, which are 
the means of the entries in Table 3 of Smith et al. (2002). The extinction 
values for May 10 - 17 1998 and March 28 - 31 1999, which were periods of abnormally 
high extinction at the USNO Flagstaff Station site, were not used when computing the 
mean coefficients. Adopting the standard deviations 
in the nightly Smith et al. extinction coefficients as the uncertainties in the values 
used here, then the error in k$_{r'}$ and k$_{i'}$ is $\pm 0.02$ mag airmass$^{-1}$. 
These introduce uncertainties of $\pm 0.01$ mag in $r'$ and $i'$ 
brightnesses. A second order extinction term was not employed 
as the coefficients are small (Smith et al. 2002).

	The zeropoints obtained from the standard star 
measurements have an internal uncertainty of $\pm 0.03$ mag filter$^{-1}$. However, the 
uncertainty in the absolute calibration of the NGC 5102 observations is larger than 
this given that the standards were not observed on the same night as NGC 5102. 
To be conservative, an uncertainty in the absolute calibration 
of $\pm 0.08$ magnitude is adopted, which is the upper limit of that suggested on 
the GMOS web pages (http://www.gemini.edu/sciops/instruments/gmos/calibration).

\section{RESULTS: A FIRST LOOK AT THE CMDs}

	The brightnesses of objects in the final GMOS images were measured 
with the point spread function (PSF)-fitting routine ALLSTAR (Stetson \& Harris 
1988). The star lists, preliminary magnitudes, and 
PSFs that were used in ALLSTAR were obtained by running 
routines in DAOPHOT (Stetson 1987). Artificial star experiments were used to assess 
completeness and estimate the uncertainties in the photometric measurements. The 
artificial stars were assigned colors and brightnesses that are appropriate 
for the locus of red stars in NGC 5102. As with the real data, an artificial star 
was considered to be recovered only if it was detected in both $r'$ and $i'$. 
The difference between the actual and measured brightness was not applied as a criterion 
for recovery, so that the effects of blending on stellar brightness could be investigated. 

	The completeness fraction varies across the field because the stellar 
density in NGC 5102 increases with decreasing R$_{GC}$. At projected radii in the 
disk plane of 3.5 kpc, the 50\% completeness limit is $i' \sim 25$ for a typical giant, 
while at 5.5 kpc the 50\% completeness limit is almost a magnitude fainter. There is also 
a systematic radial trend in the difference between the actual and measured $i'$ 
brightness of artificial stars, $\Delta i'$, that is 
another consequence of crowding. As might be anticipated from the 50\% completeness 
limits, this is not significant near $i' = 24$, which is the approximate brightness of the 
RGB-tip for metal-poor stars in NGC 5102, when R$_{GC} > 3.5$ kpc. However, 
$\Delta i'$ becomes larger at progressively fainter magnitudes. At $i' = 25$,
which is the approximate brightness of the RGB-tip for the most metal-rich stars in NGC 
5102, $\Delta i' = 0.4$ magnitudes when R$_{GC} = 3.5$ kpc, although 
$\Delta i'$ drops to 0.1 magnitudes for stars with $i' = 25$ when R$_{GC} = 5.5$ kpc. It 
should be noted that colors tend not to be as affected as single 
filter brightness measurements, as interloping objects that are 
bright enough to affect photometry in one bandpass tend to also be significant 
nuisances at neighboring wavelengths. In summary, the artificial star experiments indicate 
that caution should be exercised when assessing the properties of metal-rich 
RGB stars in the inner few kpc of NGC 5102.

	The sources detected in the GMOS images were sorted into 1 kpc wide annuli. 
The placement of objects in these annuli assumes a distance modulus of 27.5 (\S 5.1) and a 
disk inclination of 64.4 degrees to the line of sight, based on the ratio of 
major and minor axes in the 2MASS Large Galaxy Atlas (Jarrett et al. 2003). 
The CMDs of the various annuli are shown in Figure 2. The areas 
sampled in these annuli are listed in Table 1.

	The faint limit of the CMDs in Figure 2 varies with R$_{GC}$, as expected due 
to the increase in stellar density towards smaller R$_{GC}$. A plume of red stars, 
which is made up of RGB stars in NGC 5102, can be traced out to R$_{GC} \sim 
10$ kpc. While stars that belong to NGC 5102 are almost certainly present when 
R$_{GC} > 10$ kpc, their number density is such that they do not stand out against 
the population of foreground stars and background galaxies. 

	Background galaxies dominate over foreground stars near the 
brightness of the RGB-tip in NGC 5102. The Robin et al. (2003) model Galaxy predicts that 
there are $\sim 60$ foreground stars with $i'$ between 23 and 24 in the area 
imaged by GMOS, with $\sim 7$ stars in the 10 -- 11 kpc interval. For comparison, 
galaxy counts compiled in Figure 2 of Smail et al. (1995) suggest that 
there are $140 - 350$ background galaxies with $i'$ between 23 and 24 in the 
GMOS field, with $\sim 16 - 41$ galaxies in the 10 -- 11 kpc annulus. There 
are 24 objects with $i'$ between 23 and 24 in the $10 - 11$ kpc CMD in Figure 2, and 
so there is excellent agreement between the predicted and observed number of 
contaminating objects. 

	The evolutionary status of stars in NGC 5102 is examined in Figure 3, 
where the CMDs of objects in three annuli are compared with isochrones. 
The CMDs for the $3-4$ kpc and $6-7$ kpc intervals demonstrate the typical 
data quality in the NGC 5102 disk, whereas the majority of sources 
in the 10 -- 11 kpc interval are background galaxies and foreground stars. 
The errorbars show the random uncertainties predicted 
from the artificial star experiments for a star with $i' = 24$ and 
$r'-i' = 0.5$. The error ellipse shrinks as R$_{GC}$ 
increases, due to the diminishing impact of crowding on photometric 
measurements as regions with progressively lower stellar density are sampled. 

	Girardi et al. (2004) transform a grid of isochrones into the SDSS photometric 
system, and some of these are compared with the NGC 5102 observations in Figure 3. The 
isochrones shown in Figure 3 have Z = 0.0001, 0.001, 0.004, and 0.008; those in the top row 
have an age of 10 Gyr, while those in the lower row have an age 
of 1 Gyr. A distance modulus $\mu_0 = 27.5$ (\S 5.1) and a foreground reddening 
A$_B = 0.237$ (Schlegel, Finkbeiner, \& Davis 1998), which corresponds to 
A$_{i'} = 0.12$ and E$_{r'-i'} = 0.04$ using the entries in Table 6 of Schlegel et al. 
(1998), have been assumed.

	Stars with distinct ages and evolutionary states can be identified in the CMDs. 
There is a significant population of sources with $r'-i' < 0$ in the $3 - 4$ kpc CMD, 
whereas this region of the 10 -- 11 kpc CMD is almost devoid of objects. 
The blue objects in the $3 - 4$ kpc CMD are bright main sequence 
stars, the properties of which are discussed in \S 4.

	It can be seen in the top panel of Figure 3 that the majority of objects in the 
NGC 5102 CMDs have colors and brightnesses that are consistent with them evolving 
on the RGB, and these stars are discussed further in \S 5.
There are also stars above the RGB-tip in Figure 3 that 
are evolving on the AGB. The AGB stars that are immediately above the RGB-tip have 
brightnesses and colors that are consistent with them belonging to an old population; 
however, there are also AGB stars that have brightnesses indicating that they belong to 
an intermediate age population, and it can be seen from the lower panel of Figure 3 
that the peak brightness of the AGB stars is consistent with an age $\leq 1$ Gyr. 
The properties of the AGB stars are discussed in \S 6.

\section{MAIN SEQUENCE STARS AND SUPERGIANTS}

\subsection{Main Sequence Stars and the Star Formation Rate}

	The HII regions in NGC 5102 mark the sites of the most recent star formation 
in the galaxy. Although HII regions have yet to be detected in the field observed with 
GMOS, there are blue objects with $r'-i' < 0$ in the CMDs of the 2 -- 3 and 3 -- 4 kpc 
intervals in Figure 2. These objects are almost certainly young main sequence stars in NGC 
5102 rather than foreground stars or background galaxies, as objects with similar 
colors and brightnesses are absent in the CMDs that sample R$_{GC} >$ 10kpc.

	The blue stellar sequence in Figure 2 peaks near $r' \sim 22$, which 
corresponds to M$_{r'} \sim -5.5$. This is consistent with the peak 
brightness of main sequence stars in the Galaxy and the LMC 
(e.g. Humphreys \& McElroy 1984). As stars at the uppermost limits of the main sequence, 
the blue stars in the western portion of the NGC 5102 disk likely have 
ages $\leq 10$ Myr. Thus, star formation has occured recently 
throughout much of the western half of the NGC 5102 disk, albeit at a very modest rate 
(see below).

	Clues to the mechanisms that drive star formation in the disk of NGC 5102 can be 
found by examining the spatial distribution of young objects. The spatial distribution of 
main sequence stars in NGC 5102 is investigated in Figure 4, where the positions of 
sources with $i' < 25$ and $r'-i' < 0$ are shown. The stellar positions are in the 
plane of the NGC 5102 disk -- i.e. they have been de-projected to account for 
the inclination of the NGC 5102 disk. While there appears to be 
a slight tendency for the main sequence stars to cluster, they are distributed 
throughout the western portion of the NGC 5102 disk, with 
no sign of spiral structure. This is not unexpected given the absence of 
spiral structure in the NGC 5102 HI distribution (van Woerden et al. 1993). 
Thus, it appears that star formation in the disk of NGC 5102 
at the present epoch is not spurred by spiral density waves. 

	Much of the HI in NGC 5102 is in a ring that has its greatest concentration between 
2 and 4 arcmin from the galaxy center (e.g. Figure 6 of van Woerden et al. 1993). The 
dashed lines in Figure 4 show the annulus that marks the 
HI ring in this portion of NGC 5102. Many of the bright main sequence stars in 
the western portion of NGC 5102 fall outside of the HI ring, suggesting that the 
large scale HI distribution in NGC 5102 is not 
an effective tracer of recent star formation.

	The density of main sequence stars in the disk of NGC 5102 is considerably lower 
than in nearby `normal' late-type spirals, and this can be quantified by investigating the 
specific frequency of blue stars. Following the convention developed to probe globular 
clusters systems (e.g. Harris \& van den Bergh 1981), the specific frequency of young stars 
is taken to be the number of blue objects per unit integrated brightness. For the 
current study, the specific frequency of main sequence stars, S$_{ms}$, is defined to 
be the number of stars with $r'-i' < 0$ and $r'$ between 22 and 24 
normalized to a system with an integrated brightness M$_K = -16$. Therefore, 
S$_{ms} = n_{ms} \times 10^{0.4(M_K + 16)}$, where n$_{ms}$ is the number of stars 
with $r'-i' < 0$ and $r'$ between 22 and 24, while M$_K$ is the integrated brightness 
in the region of NGC 5102 being studied. There are 44 stars with 
R$_{GC}$ between 2 and 5 kpc in the GMOS field that have photometric properties that 
suggest they are main sequence stars. Using the total $K-$band brightness in this radial 
interval measured from the 2MASS image of NGC 5102 (Jarrett et al. 2003), then S$_{ms} = 
9$. For comparison, the star counts given by Davidge (2007b), coupled with the 
$K-$band surface photometry from Jarrett et al. (2003), indicate that 
S$_{ms} = 140$ in the disk of NGC 2403. Therefore, when normalized to the same integrated 
$K-$band brightness the number density of main sequence stars in NGC 5102 is 
only $\sim 7\%$ that in NGC 2403.

	The recent SFR in NGC 5102 can be estimated using NGC 2403 as a benchmark. 
The total SFR in a spiral galaxy can be computed using diagnostics such as the strength 
of H$\alpha$ emission, which probes star formation in regions of low obscuration, and 
infrared flux, which probes star formation in heavily obscured areas (e.g. 
P\'{e}rez-Gonz\'{a}lez et al. 2006). Using the H$\alpha$ and FIR fluxes of NGC 2403 
given in Table 4 of Bell \& Kennicutt (2001) and the SFR calibrations given by 
Kennicutt (1998), then the unobscured SFR in NGC 2403 is $\sim 0.4$ M$_{\odot}$ 
year$^{-1}$, while the obscured SFR is 0.3 M$_{\odot}$ year$^{-1}$. Given that (1) the 
total M$_K$ magnitude of NGC 5102 is $\sim 0.8$ magnitudes fainter than NGC 
2403 (Jarrett et al. 2003), and (2) the density of young 
stars in NGC 5102 is 7\% that in NGC 2403, then the present day unobscured 
SFR in NGC 5102 is $\sim 0.013$ M$_{\odot}$ year$^{-1}$. As for the obscured SFR in NGC 
5102, the total far infrared flux in NGC 5102 is $\sim 2\%$ that in NGC 2403 (Rice et al. 
1988), and so the obscured SFR in NGC 5102 is $\sim 0.006$ M$_{\odot}$ year$^{1}$. 
Therefore, the total SFR in NGC 5201 is $\sim 0.02$ M$_{\odot}$ year$^{-1}$.

\subsection{Red Supergiants}

	Red supergiants (RSGs) populate a near-vertical finger of stars in the $2 - 3$ kpc 
CMD in Figure 2. The RSG sequence runs from $i' = 22.5$ to $i' = 21.4$, with $r'-i' 
\sim 0.3$. A similar feature is seen in the $(i', r'-i')$ CMDs of disk stars in
NGC 2403 (Davidge 2007b) and NGC 247 (Davidge 2006). It should be emphasized that 
despite the low number density of young stars throughout the disk of NGC 5102, 
the RSG sequence in the $2 - 3$ kpc CMD is a statistically 
real feature. Indeed, whereas there are 13 sources with 
$i'$ between 22.5 and 21.4 and $r'-i'$ between 0.2 and 0.5 in the 2 -- 3 kpc 
CMD, the number of similar objects in the 10 -- 11 kpc interval indicates that only 
$\sim 1$ of the candidate RSGs in the $2 - 3$ kpc CMD is probably a foreground star 
or background galaxy.

	The color and shape of the RSG locus on visible wavelength CMDs is sensitive to 
metallicity, and so it is possible to estimate the metallicity of 
RSGs in NGC 5102. This is demonstrated in Figure 5, where models 
with Z = 0.008 and Z = 0.019 from Girardi et al. (2004) are compared with the 
2 -- 3 kpc CMD. The RSG locus is much better matched by the Z = 0.008 models 
than the Z = 0.019 models.

	The metallicity of RSGs provides constraints on the metallicities 
of older stars in the disk of NGC 5102. Given that the youngest 
stars in a galaxy will probably form from interstellar material that has been subject to 
cumulative enrichment from past generations of stars, then 
RGB stars with metallicities that are higher than those of the RSGs should be rare 
in the disk of NGC 5102. It can then be anticipated that the metallicity distribution of 
RGB stars in NGC 5102 will truncate near [M/H] $\sim -0.3$, and this is seen to be the 
case (\S 5.2).

	The spatial distribution of RSGs in the western disk 
of NGC 5102 is examined in the right hand panel of Figure 4. 
The objects identified as RSGs in this figure have the same photometric properties 
as marked in Figure 5. It is evident that -- as was found for main 
sequence stars -- there are a number of RSGs that fall outside the HI ring boundaries.

\section{THE RED GIANT BRANCH}

\subsection{An RGB-tip Distance Modulus}

	The distance to NGC 5102 can be measured from the brightness of the RGB-tip. 
A complication is that the RGB sequence in the disk of NGC 5102 is broad, and 
the isochrones in the top row of Figure 3 indicate that 
the RGB stars in this part of NGC 5102 span metallicities from Z = 0.0001 to 
at least Z = 0.008. Such a broad metallicity spread complicates efforts to measure 
the brightness of the RGB-tip at visible and red wavelengths, since the upper RGB 
bends over when [Fe/H] $> -0.7$ (e.g. Da Costa \& Armandroff 1990; Ortolani, 
Barbuy, \& Bica 1991), which in turn blurs the signature RGB-tip
discontinuity in the luminosity function (LF). 

	Age is another parameter that can broaden the RGB and affect the ability to 
detect the RGB-tip. At a given metallicity, younger stars have bluer colors than 
older stars, and the 1 Gyr AGB sequences in the bottom row of Figure 3 pass through a 
part of the CMD that is also occupied by metal-poor RGB stars. The presence of 
these intermediate age AGB stars will dilute the RGB-tip discontinuity.

	The impact of a broad metallicity dispersion can be reduced by restricting 
the RGB sample to stars with colors that are appropriate for Z $\leq 0.002$. For the 
current study only stars with $r'-i'$ between 0.3 and 0.6 are used to determine 
the RGB-tip brightness. While the isochrones shown in the 
lower panel of Figure 3 indicate that intermediate age stars may be present in this 
color interval, the removal of these objects with the available data is problematic. 
In any event, while AGB stars with this age may reduce the amplitude of the RGB-tip 
discontinuity, they are not expected to skew the brightness 
estimated for the RGB-tip. The sample was further culled so that only stars 
with R$_{GC} > 4$ kpc are considered. This was done to reduce the impact of photometric 
errors on the RGB-tip measurement. 

	The linear and logarithmic LFs of stars with R$_{GC}$ between 4 and 7 kpc and
having $r'-i'$ between 0.3 and 0.6 are shown in the upper two panels of Figure 6. 
The result of convolving the linear LF with a three point Sobel edge-detection kernel 
is shown in the bottom panel. There is a peak in the filtered LF at 
$i' = 24.15 \pm 0.13$, which is attributed to the RGB-tip. The quoted uncertainty in the 
RGB-tip brightness is the result of adding in quadrature the error due to binning, which is 
estimated to be $\pm 0.10$ based on the width of the peak in the lower panel of Figure 6, 
with the error in the photometric calibration, which is $\pm 0.08$ magnitude (\S 2).

	The conventional brightness calibration of the RGB-tip is in the $I_{KC}$ filter 
(e.g. Lee, Freedman, \& Madore 1993). To use this calibration, the GMOS $i'$ magnitude 
of the RGB-tip was transformed into an I$_{KC}$ magnitude using the transformation 
equations in Table 7 of Smith et al. (2002). If the $I-$band RGB-tip brightness is 
M$_{I}^{RGBT} = -4$ (Lee et al. 1993) and A$_B = 0.237$, then a 
distance modulus of $27.51 \pm 0.13$ is found for NGC 5102.

	The estimated uncertainty in the distance modulus 
is a lower limit, as it does not account for uncertainties in the transformation 
into I$_{KC}$ magnitudes. However, there are indications that 
this transformation does not introduce large errors. One way to 
determine if the transformation process has seriously skewed the 
distance modulus is to compute an independent distance modulus from the untransformed 
$i'$ RGB-tip magnitude. As it turns out, a distance modulus of 27.50 is computed based on 
the RGB-tip M$_{i'}$ brightness predicted by the 10 Gyr Z=0.001 Girardi et al. (2004) 
isochrone. The good agreement between this distance modulus and that 
computed from the Lee at el. (1993) calibration in the previous paragraph argues 
that the transformation errors are probably not large.

	An important caveat in the distance estimate is that the RGB stars are 
assumed to be like those in globular clusters, and this is not an ideal assumption for NGC 
5102, where there is a large intermediate age population. Salaris \& Girardi (2005) 
investigate the impact of intermediate age populations on RGB-tip distance estimates, 
and find that the distance modulus may be in error by $\sim 0.1$ magnitude. 
Salaris \& Girardi (2005) note that there are significant uncertainties when correcting the 
RGB-tip distance modulus for intermediate age populations, due in part to uncertainties in 
the bolometric corrections of cool stars. In addition, the distance modulus correction 
also depends on the stellar content. Assigning an uncertainty of $\pm 
0.1$ magnitude for population effects, which roughly brackets the 
range of corrections computed by Salaris \& Girardi (2005) when modelling the relative 
distances of the Large and Small Magellanic Clouds with respect to LGS3, then the final 
distance modulus for NGC 5102 based on the GMOS data is $27.51 \pm 0.16$.

	The distance modulus found from the GMOS data does not differ significantly 
from that computed by Karachentsev et al. (2002), who concluded that $\mu_0 = 
27.66 \pm 0.25$ based on observations of a different part of 
NGC 5102 than was observed with GMOS. The distance modulus computed here is 
adopted for the remainder of the study simply because it is internally consistent with 
the GMOS data. Given that the two RGB-tip distance moduli are not significantly different, 
we otherwise do not prefer one over the other.

\subsection{The Metallicities of RGB Stars}

	Line blanketing affects the $r'$ and $i'$ brightnesses of evolved stars with 
Z $\geq 0.004$, in that evolutionary sequences for metal-rich and moderately metal-rich 
systems bend over on $(i', r'-i')$ CMDs near the bright end; consequently, the 
most evolved stars on the RGB and AGB may have fainter observed magnitudes than stars that 
are less evolved. The impact of line blanketing is greatly reduced at wavelengths longward 
of $1\mu$m, and a survey of NGC 5102 in $J, H,$ and $K$ will allow for a more 
complete census of the evolved red stellar content of this galaxy.
In the absence of such data, the impact of line blanketing on the morphology of 
the metal-rich portions of the RGB can be reduced by 
considering the bolometric magnitudes of individual stars. 

	A relation between the $i'$ bolometric 
correction, BC$_{i'}$, and $r'-i'$ was generated from the Girardi et al. (2004) 
isochrones. The following three-part relation between BC$_{i'}$ and $r'-i'$ was found to 
match the models with Z $\geq 0.001$ to within $\pm 0.1$ magnitudes, independent of 
metallicity and age:

\begin{displaymath}
BC_{i'} = -1.10 (r'-i') + 0.50 \hspace{1.0cm} (0.20 < r'-i' < 0.70)
\end{displaymath}

\begin{displaymath}
BC_{i'} = -0.58 (r'-i') + 0.14 \hspace{1.0cm} (0.70 < r'-i' < 1.15)
\end{displaymath}

\begin{displaymath}
BC_{i'} = -1.48 (r'-i') + 1.17 \hspace{1.0cm} (1.15 < r'-i' < 2.00)
\end{displaymath}

\noindent{The} (M$_{bol}$, $r'-i'$) CMDs for the 3 -- 4 kpc, 5 -- 6 kpc, and 11 -- 12 kpc 
intervals are shown in Figure 7. Isochrones from Girardi et al. (2004) for Z = 0.0001, 
0.001, 0.004, and 0.008 are also shown.

	The metallicity distribution function (MDF) of RGB stars probes the chemical 
enrichment history of galaxies. The metallicities of individual RGB stars can be 
estimated from their colors, and the MDF can be computed from these values. 
Photometric metallicities were computed for stars with M$_{bol}$ between --2.9 
and --3.1 by interpolating between the 10 Gyr Girardi et al. (2004) isochrones. This 
M$_{bol}$ magnitude range was selected because it samples the RGB at a point where the 
data are largely complete throughout most of the area 
that was observed with GMOS. The MDFs constructed from these data are 
shown in Figure 8. The MDFs have been corrected for foreground 
star and background galaxy contamination by subtracting 
the MDF of objects in the 10 -- 12 kpc interval. 

	The MDFs are dominated by objects with [M/H] between 
--1.3 and --0.1. Caution should be exercised when considering the metal-rich end of each 
MDF, as this is where incompleteness will be greatest. The MDFs in Figure 8 
are resticted to R$_{GC} > 4$ kpc, as the artificial star experiments suggest 
that incompleteness is not an issue for metal-rich stars at these galactocentric 
distances. In fact, there are two indirect indications that incompleteness 
does not influence the overall characteristics of the MDFs in Figure 8. 
First, it was shown in \S 4 that the RSGs in NGC 5102 have [Fe/H] $\sim -0.3$, and 
it is unlikely that the (older) RGB disk stars will be more metal-rich than this. 
Therefore, the absence of a large number of RGB stars with [M/H] $\geq -0.1$ in Figure 8 is 
to be expected. Second, the degree of incompleteness at a given magnitude and color 
will be greater in regions of higher stellar density, and so it might be expected that the 
metal-rich ends of MDFs at small R$_{GC}$ will truncate at 
lower metallicities that at large R$_{GC}$. That the MDFs in Figure 8
appear not to change near the metal-rich end over a large range of R$_{GC}$ suggests 
that incompleteness is likely not a major issue in defining the MDFs. 

	Crowding is another source of concern when considering the 
MDFs. If two stars of the same brightness fall in the same angular resolution 
element then they will appear as a single object that is 0.75 mag brighter than 
the progenitors. This will have the greatest impact among metal-rich RGB stars, where the 
RGB-tip is fainter in $i'$ than for metal-poor stars. The incidence of blending can be 
estimated empirically by considering the number density of objects with $i'$ between 
25.5 and 26.0, which is the magnitude range where pairs of stars will blend to form 
objects with brightnesses that are comparable to those of metal-rich RGB-tip stars. 
Accounting for incompleteness, there are 3240 stars with $i'$ between 25.5 and 26.0 
in the 4 -- 6 kpc interval, and 630 such stars in the 6 -- 8 kpc interval. 
Assuming that each resolution element has a radius that is one half of the FWHM, 
then there will be $\sim 160$ two-star blends in the 4 -- 6 kpc interval and 
only 6 such blends in the 6 -- 8 kpc interval.

	The number of blends predicted in the previous paragraph are conservative upper 
limits. First, stars that are separated by half of the seeing disk radius will form a 
composite object that is noticeably elongated, and hence will be rejected by DAOPHOT. 
This is an important consideration, as the predicted number of blends is critically 
sensitive to the size of the resolution element. 
For example, if the resolution element is assumed 
to have a radius that is one quarter that of the seeing disk, then the number of blends 
expected in the 4 - 6 kpc interval plummets to 10. Second, the number counts used to 
estimate the blending rate span the entire range of colors, and so not all blends will 
have colors that are consistent with those of metal-rich stars.

	The issue of blending at the metal-rich end 
of the MDF of stars in the 4 -- 6 kpc interval notwithstanding, 
the overall appearance of the MDFs in Figure 8 suggests that the peak [M/H] may 
shift to more metal-poor values as R$_{GC}$ increases. Such a trend is not 
unexpected given the evidence for radial metallicity gradients in the disks of many 
nearby spiral galaxies based on the measurements of line strengths in 
HII regions. Still, a Kolomogorov-Smirnov test indicates that 
the 6 -- 8 and 4 -- 6 kpc MDFs differ at well below the 
90\% confidence level -- i.e. they are not significantly different. However, 
the 8 - 10 and 4 - 6 kpc MDFs differ at roughly the 95\% confidence level. Thus, there is 
tentative evidence for a metallicity gradient in the RGB content in the NGC 5102 
disk. Larger area surveys of the outer regions of NGC 5102 will be required to 
obtain the stellar sample sizes that will allow for a more rigourous test for a 
metallicity gradient.

	The MDFs in Figure 8 are contaminated by AGB stars that belong to an 
intermediate age population (\S 6). It can be seen from Figure 7 that AGB stars with an 
age of 1 Gyr and M$_{bol} \sim -3$ will masquerade as RGB stars with [Fe/H] $\leq -1$, 
and so interloping AGB stars will skew the true MDF of old RGB stars 
to lower values. It is thus worth noting that the majority of stars in the 
MDFs have [Fe/H] $\geq -1$, indicating that intermediate age AGB stars
likely do not have a large impact on the observed MDFs. 

\section{THE ASYMPTOTIC GIANT BRANCH}

	Comparisons with isochrones suggest that (1) the brightest of the AGB stars in 
the western disk of NGC 5102 have an age $\leq 1$ Gyr, and (2) the AGB stars are 
moderately metal-rich, with the majority falling to the right of the Z = 0.004 
isochrones in the lower panel of Figure 7. This second result is not unexpected given 
that the RGB MDFs peak near Z = 0.004 (\S 5.2); the brightest AGB stars 
are younger than the RGB stars, and so likely formed from more metal-enriched 
gas than the majority of RGB stars.

	The M$_{bol}$ LFs of stars with $r'-i' > 0.2$ are compared in Figure 9. Relations 
between age and peak AGB magnitude from the Girardi et al. (2004) models are also 
indicated at the top of this figure. There are three distinct luminosity regimes 
that are considered here:

\begin{enumerate}

\item M$_{bol} > -3.5$. Stars evolving on the RGB dominate in this luminosity interval 
in systems with ages in excess of $\sim 2$ Gyr. A statistically significant 
population of stars with this brightness is seen out to R$_{GC} \sim  
10$ kpc in the LFs.

\item M$_{bol}$ between --3.5 and --4.5. This magnitude range is dominated by
AGB stars that belong to old stellar systems. Like RGB stars, an old AGB 
component can be traced in significant numbers out to R$_{GC} \sim 10$ kpc in NGC 5102.

\item M$_{bol}$ between -4.5 and -6.5. This magnitude interval 
contains intermediate age AGB stars. Stars in this luminosity range
can be traced in significant numbers out to R$_{GC} \sim 8$ kpc, 
indicating that there was star formation at intermediate epochs 
throughout much of the disk of NGC 5102. The stars in this luminosity range 
are the focus of attention in the remainder of this section. 

\end{enumerate}

	There is no obvious discontinuity near M$_{bol} \sim -4.5$ 
in Figure 9, as might be expected if the intermediate age 
stars formed in a discrete, modestly-sized star-forming episode. Rather, the continuous 
nature of the LF near M$_{bol} \sim -4.5$ hints that the brightest AGB stars belong to 
a population that contributes significantly to the stellar mass of the disk. 
The fraction of the stellar disk mass that formed during intermediate epochs 
can be estimated by comparing the LFs in Figure 9 with fiducial LFs of systems with known 
age and mass. Davidge (2007a) constructed an AGB $i'$ LF from observations 
of stars in the six well-characterised LMC clusters 
that were observed by Ferraro et al. (2004). The clusters have a 
luminosity-weighted age $\sim 0.7$ Gyr. These same data are used here to construct a 
reference LF that is compared with the NGC 5102 M$_{bol}$ LF. 

	Following Davidge (2007a), the model LFs are restricted to two components, 
consisting of an old and intermediate age population that are combined 
in various mass ratios. The LF of the old component is taken from observations 
of the globular cluster 47 Tuc. This cluster is an ideal LF template for 
the present work because it has a metallicity that is comparable to that of the majority 
of RGB stars in NGC 5102, and it is known to be old. The observations of 47 Tuc 
are unpublished, and were obtained with the CIRIM imager on the 
CTIO 1.5 meter telescope during an observing run in July 1996. The central $2.6 
\times 2.6$ arcmin$^2$ of the cluster was observed through $J$ and $Ks$ filters, with a 
total exposure time of 12 sec filter$^{-1}$. The data were reduced using a standard 
pipeline for near-infrared imaging (e.g. Davidge 2000). Stellar brightnesses were 
measured with ALLSTAR (Stetson \& Harris 1988).

	Bolometric magnitudes for individual stars in 
the LMC clusters were calculated using a relation 
between the $K-$band bolometric correction, BC$_K$, and $J-K$ that was generated from 
the 0.5 Gyr Z = 0.004 models of Girardi et al. (2004). The bolometric magnitudes of stars 
in 47 Tuc were calculated using the bolometric corrections discussed in \S 5.2. 
The masses of the old and intermediate aged template clusters were estimated from the 
total $K-$band brightnesses of the clusters using M/L ratios from Maraston (2005). 
Photometric measurements made by Pessev et al. (2006) and 
Persson et al. (1983) indicate that the LMC clusters have a total brightness $K = 7.3$, 
which corresponds to M$_K = -11.3$ assuming an LMC distance modulus $\mu_0 = 
18.5$. Images from the 2MASS survey indicate that the area observed in 47 Tuc has a 
total brightness M$_K = -10.8$. 

	Selected model LFs are shown in Figure 10, where f$_{int}$ is the fractional 
mass of the intermediate age component; the fractional mass of the old population is 
then $1 - f_{int}$. The model M$_{bol}$ LF flattens with increasing f$_{int}$. 
The models are most sensitive to changes in f$_{int}$ 
when f$_{int} \leq 0.2$, since at larger values of f$_{int}$ the intermediate age 
population dominates the light output by virtue of its higher M/L ratio.

	Model LFs are compared with the LF of stars in NGC 5102 with R$_{GC}$ between 3 
and 5 kpc in Figure 11. This radial interval was selected to avoid the incompleteness 
issues that affect the data when R$_{GC} < 3$ kpc, while the 2MASS data do not 
go deep enough to allow surface photometry to be done when R$_{GC} > 5$ kpc. 
The models have been shifted to show the number of stars expected in a system with 
an integrated brightness like that in the 3 -- 5 kpc interval in 
NGC 5102, which was measured from the $K-$band image of this 
galaxy in the 2MASS Large Galaxy Atlas (Jarrett et al. 2003). It can be seen that 
when f$_{int} < 0.2$ the model LFs are too steep when compared with the observations.
However, the number counts of stars in NGC 5102 are matched by the models with 
f$_{int} \geq$ 0.2. Therefore, a significant fraction of the 
stellar mass in the disk of NGC 5102 formed during intermediate epochs; 
elevated SFRs were not restricted to the bulge of NGC 5102 within the past 
$\sim 1$ Gyr, but also occured throughout much of the disk.

	Because the model LFs do not change by large amounts 
when f$_{int} > 0.2$, it is difficult to pin down f$_{int}$ solely from the LFs. 
However, other information provides independent constraints. For example, f$_{int} 
<< 1.0$ given that a large RGB component is present. 
Additional evidence that supports f$_{int} \sim 0.2$ comes from the number of 
metal-poor stars in the RGB MDFs, and this is discussed further in \S 7.2.

\section{SUMMARY \& DISCUSSION}

	Deep $r'$ and $i'$ images obtained with GMOS on Gemini South have been used 
to probe the stellar content in a field that samples the western side of the 
Centaurus Group S0 galaxy NGC 5102. With a distance of only 3.2 Mpc, NGC 5102 
is the nearest S0 galaxy, and so is a potential Rosetta Stone for understanding the 
origins and evolutionary state of this galaxy type. The stellar content of NGC 5102 
points to a highly interesting star-forming history, and this is discussed below.
A caveat is that NGC 5102 has a number of features that make it an unusual S0. 

	While there is some uncertainty in the distance 
modulus of NGC 5102 (\S 5.1), this will not have a major impact on the 
conclusions. For example, if the slightly larger distance modulus computed 
by Karachentsev et al. (2002) is adopted instead of the value computed here 
then the age of the youngest stars in NGC 5102 will be lowered by only 0.05 
-- 0.1 dex. The fraction of the galaxy that formed 
during intermediate epochs and the SFR estimated for this 
time will increase by modest amounts. As for the impact 
on older stars, the MDF estimated from RGB stars will be shifted to 
lower values by $\sim 0.1$ dex at the metal-rich end.

\subsection{The Star Forming History of NGC 5102 From the Present Day to 1 Gyr in the Past}

	There is on-going star formation in the disk of NGC 5102, albeit at a 
subdued level. The most obvious signature of this activity is the 
modest population of HII regions that are clustered in the eastern and southern areas 
of the galaxy. A new result is that the GMOS observations reveal that recent star 
formation has occured in other parts of the NGC 5102 disk, as bright main 
sequence stars, which probe star-forming activity up to $\sim 10$ Myr in the past, 
are seen throughout the western disk of NGC 5102. The spatial distribution of these 
objects suggests that star formation during the past $\sim 10$ Myr was not spatially 
concentrated, but occured at various locations throughout this part of NGC 5102.
Given the absence of obvious spiral structure (van Woerden et al. 1993), the spatial 
concentration of HII regions and the diffuse distribution of bright main sequence stars 
suggest that star formation at the present day in the disk of NGC 5102 is not 
spurred by spiral density waves; perhaps star formation is spatially propogating 
from many seed sights.

	The global SFR in NGC 5102 has not changed significantly over the past 10 Myr. 
Indeed, the density of main sequence stars per $K-$band surface brightness in NGC 5102 
indicates that the SFR has been $\sim 3\%$ that in NGC 2403 during the past 
$\sim 10$ Myr (\S 4.1). A comparison of the number of HII regions in these galaxies 
can be used to judge their relative SFRs over an even shorter timescale (a few Myr). Sivan 
et al. (1993) find 366 HII regions in NGC 2403, while Hodge \& Kennicutt (1983) identify 
605 HII regions. For comparison, there are only 7 HII regions known in 
the main body of NGC 5102 (McMillan et al. 1994), although this is probably a lower 
limit as not all of the NGC 5102 disk was surveyed by McMillan et al. (1994). 
Using integrated K-band brightnesses from Jarrett et al. (2003), then the number of 
HII regions per unit $K-$band brightness in NGC 5102 is $\sim 
0.02 - 0.04 \times$ that in NGC 2403. This agrees with the spatial densities of the 
longer-lived bright main sequence stars.

	The SFR in NGC 5102 was much higher a few 100 Myr 
in the past. In \S 6, it is concluded that at least 20\% 
of the total stellar disk mass in NGC 5201 formed within the past $\sim 1$ Gyr. 
Kraft et al. (2005) argue that a large fraction of the stellar mass in the 
central regions of NGC 5102 also formed during intermediate epochs. 
It thus appears that the elevated levels of star 
formation within the past Gyr were not restricted to the central spheroid 
of NGC 5102, but occured throughout much of the galaxy.

	The detection of an intermediate age population in a normal spiral galaxy should 
not in itself be surprising, as star formation proceeds in a roughly continuous manner 
in these systems. Consider the Milky-Way, where the global SFR is $\sim 3$ M$_{\odot}$ 
year$^{-1}$, and likely has not varied greatly with time over the past few Gyr (e.g. 
Naab \& Ostriker 2006; Rana 1991). Given a disk mass of $7 \times 10^{10}$ 
M$_{\odot}$ (van der Kruit 1986), then $\sim 4\%$ of the stellar mass in the 
Galactic disk formed between 0.2 and 1.0 Gyr in the past. The model LFs 
discussed in \S 6 indicate that an external 
viewer of the Galaxy would detect a prominent bright AGB sequence. 
This being said, the fractional mass of the disk of NGC 5102 that formed during 
intermediate epochs suggests that the SFR at those times exceeded that of a normal 
spiral galaxy, and certainly exceeded the present day SFR in NGC 5102. 
Given that at least 20\% of the stellar disk mass in NGC 5102 appears to have 
formed sometime in the past 1 Gyr then the SFR per projected area in the NGC 
5102 disk must have been at least $5 \times$ that in the Milky-Way disk.

	A lower limit to the SFR in NGC 5102 during intermediate 
epochs can be estimated from the available data. 
van Woerden et al. (1993) find a total mass of $1.2 \times 10^{10}$ M$_{\odot}$ within 6 
kpc of the center of NGC 5102. Adopting a mean dark matter density of $5 \times 
10^{-3}$ M$_{\odot}$ pc$^{-3}$ (Puche \& Carignan 1991), then the stellar mass in NGC 
5102 is $7 \times 10^{9}$ M$_{\odot}$. This stellar mass is not unreasonable, as it is 
within a factor of two of what is computed from the total $K-$band brightness of 
NGC 5102 from Jarrett et al. (2003) assuming $M/L_K = 1$. The total 
stellar mass that formed during intermediate epochs is then at least $1.4 \times 10^{9}$ 
M$_{\odot}$. Star formation in starburst systems occurs 
over an extended period of time. Studies of distant luminous infrared 
galaxies suggest a typical time scale of 0.1 Gyr (Marcillac 
et al. 2006), while observations of closer starburst galaxies 
like M82 suggest timescales of $\sim 0.3$ Gyr (Mayya et al. 2006) 
to 1 Gyr (e.g. Smith et al. 2006). Conservatively adopting the longer 
of these timescales, then the SFR in NGC 5102 during 
intermediate epochs was then at least 1.4 M$_{\odot}$ year$^{-1}$. 

\subsection{The Early Evolution of NGC 5102}

	While a significant fraction of the NGC 5102 disk formed during 
intermediate epochs, there is also a large old stellar substrate. 
Evidence that an RGB is present, and that the 
disk is not comprised entirely of intermediate age stars, comes from the 
red envelope of objects with M$_{bol} > -4$ in the $(M_{bol}, r'-i')$ CMD, which  
can only be moderately metal-rich RGB stars. There is also a clean RGB-tip 
discontinuity in the $i'$ LF (\S 5.1).

	Clues into the early evolution of NGC 5102 can be gleaned from 
studies of the oldest objects. Consider the globular cluster content of NGC 5102. 
The specific frequency of globular clusters in NGC 5102 is low 
when compared with other S0s (Kraft et al. 2005), but 
is not atypical for late-type spiral galaxies. 
The formation of globular clusters are associated with the 
early violent epochs of galaxy formation, that produce pressure-supported spheroids. 
The low specific globular cluster frequency argues against a violent beginning for 
NGC 5102, such as a major merger of gas rich proto-galaxies. 

	The RGB stars discussed in \S 5 are among the brightest individual old stars in 
NGC 5102. The MDFs constructed from the RGB stars suggest that the vast majority of 
old stars in the disk of NGC 5102 have metallicities between [Fe/H] $= -1.3$ and 
--0.1. That there are old stars with moderately high metallicities  
hints at much star-forming activity in NGC 5102 at epochs more 
than 3 Gyr in the past. Comparisons with other galaxies are of 
interest in this regard. The width and peak metallicity of the NGC 5102 MDF is 
comparable to the MDF in the inner disk of the LMC constructed by Cole, Smecker-Hane, 
\& Gallagher (2000). However, the NGC 5102 MDF is skewed to lower metallicities 
when compared with the MDF in the Solar neighborhood (e.g. Wyse \& Gilmore 1995) and 
the disk of M31 (Bellazzini et al. 2003), both of which contain a significant number of 
objects with [M/H] $> -0.1$.

	Line blanketing can introduce a bias against the detection of the most metal-rich 
stars if they fall in areas of the CMD with low photometric completeness. The MDFs were 
constructed at R$_{GC}$ where artificial star experiments suggest that incompleteness 
and blending should not be a problem. It is also encouraging that the 
RSGs in the disk of NGC 5102 have colors that 
are indicative of [M/H] $\sim -0.3$. As relatively young stars, the RSGs 
should be among the most metal-rich stars in the disk of 
NGC 5102, and it is unlikely that there will be a large population of 
disk RGB stars with metallicities higher than this value. The metal-rich end 
of the RGB MDF is consistent with this expectation. The 
affects of line blanketing are greatly reduced in the near-infrared, and deep 
$JHK$ imaging of the disk of NGC 5102 will provide a means of determining if a substantial 
population of RGB stars with near-solar metallicity are present. 

	Contamination by intermediate age stars is another potential source of 
uncertainty in the MDF. The MDF was measured using stars with M$_{bol} \sim -3$. At this 
magnitude, AGB stars from intermediate age populations have colors that are the same 
as those of metal-poor RGB stars, and the presence of unidentified AGB interlopers will 
skew the MDF to lower values. In fact, some of the stars identified as metal-poor 
RGB stars in \S 5 may actually be intermediate age AGB stars. It can be seen 
from Figure 7 that stars with an age of 1 Gyr and Z = 0.008 have 
$r'-i'$ colors at M$_{bol} = -3$ that are similar to those of RGB stars with 
[M/H] $\sim -1.0$. Now, the model LFs predict that if f$_{int} = 0.2$ 
then $\sim 40\%$ of the stars with M$_{bol} = -3$ will come from the 0.7 Gyr 
population. For comparison, $\sim 30\%$ of the stars in the MDFs of the  
4 -- 6 and 6 -- 8 kpc intervals have [M/H] $\leq -1$. Thus, the number of stars 
with [M/H] $\leq -0.9$ is not greatly different from that expected if (1) 20\% of the 
disk mass comes from intermediate age stars, and (2) the number of RGB stars in the 
disk of NGC 5102 with [M/H] $\leq -0.9$ is modest. In summary, accounting for intermediate 
age stars will truncate the RGB MDF near [M/H] $\sim -0.9$, rather than at [M/H] 
$\sim -1.3$. The peak in the MDF of the 4 -- 6 kpc interval 
at [M/H] $= -0.6$ is not affected by intermediate age stars; the conclusion that the 
majority of disk giants in NGC 5102 are moderately metal-rich is thus robust.

	We close the discussion of the old stellar content of NGC 5102 by noting that 
the old disk of this galaxy is spatially extended when compared with 
late-type spirals. In \S 3 it is shown that RGB stars can be traced out to R$_{GC} 
\sim 10$ kpc along the disk plane in NGC 5102. This is a lower limit to the actual extent 
of the stellar disk, as it indicates only the point at which the number of RGB stars 
can be detected in the presence of contamination from background galaxies and 
foreground stars. With a distance-corrected exponential scale length of 0.7 kpc 
(Freeman 1970), then the disk of NGC 5102 is traced out to $\sim 14$ scale lengths. 
This exceeds what is measured in many late-type spirals, such as NGC 247 
(Davidge 2006), NGC 2403 (Davidge 2007b), and even NGC 300 (Bland-Hawthorn et al. 2005).

\subsection{NGC 5102 as a Post-Starburst Galaxy}

	There are hints that NGC 5102 may have been a late-type disk galaxy earlier 
in its life. The low specific frequency of globular clusters 
is consistent with that of a late-type spiral galaxy, rather than of an early-type system 
(Kraft et al. 2005), while the MDF of RGB stars in the disks of NGC 5102 and the LMC 
are similar. In addition, while there are no signs of spiral structure in the 
distributions of young stars or HI, the radial distribution of HI in NGC 
5102 more closely follows that in spirals than classical S0s, 
albeit with a lower mass density (van Woerden et al. 1993). 

	A SFR like that which occured in NGC 5102 during intermediate epochs may have been 
high enough to have a significant impact on the observational 
properties of a normal late-type disk galaxy. Heckman (2002) finds that outflows occur 
when the density of star formation exceeds $\sim 0.1$ solar masses year$^{-1}$ kpc$^{-2}$. 
With an effective radius of 75.6 arcsec (Jarrett et al. 2003), which 
corresponds to a linear distance of 1.2 kpc, then a SFR of $\sim 1$ 
M$_{\odot}$ year$^{-1}$ is required to fuel an outflow in NGC 5102. This falls 
below the lower limit to the SFR during intermediate epochs computed in \S 7.1 (1.4 
M$_{\odot}$ year$^{-1}$), and raises the possibility that NGC 5102 may have lost 
much of its gas and dust in the not too distant past. The result 
would be a disk-dominated galaxy with a depleted ISM and a low SFR at the present day.

	The events that lead to elevated SFRs in the nucleus and bulge of 
NGC 5102 during the past $\sim 1$ Gyr are a matter of 
speculation. There are no morphological features that support a 
merger-driven origin for the elevated SFRs, and NGC 5102 is in an isolated 
environment (van Woerden et al. 1993). Still, the distance estimates listed in Table 2 of 
Karachentsev et al. (2007) place the barred spiral galaxy ESO383--G087, which appears 
to be the closest known neighbor, at a distance of $\sim 0.3 - 0.4$ Mpc, 
while Cen A is at a distance of $\sim 0.6$ Mpc. While 
NGC 5102 is currently isolated, a triggering encounter may still have occured 
$\sim 1$ Gyr in the past. Indeed, the distance between NGC 5102 and ESO383-G087 could be 
traversed within 1 Gyr if the velocity difference between these galaxies were 300 -- 400 km 
sec$^{-1}$.

	There may be galaxies in other nearby groups that are related to NGC 5102 in 
an evolutionary context. The high SFR estimated for NGC 5102 during 
intermediate epochs is reminiscent of what is seen in M82. The SFR of M82 based on the 
FIR flux is $\sim 6$ M$_{\odot}$ year$^{-1}$, and O'Connell \& Mangano (1978) suggest 
that M82 was likely an Sc/Irr galaxy prior to interacting with M81. 
Using a variety of observational constraints, including 
chemical abundances and the depth of Balmer absorption lines, Mayya et al. (2006) 
conclude that a large fraction of the M82 disk formed over a period of only a few 
hundred Myr, terminating $\sim 0.5$ Gyr in the past. The star formation is powering a mass 
outflow in M82 (e.g. Heckman, Armus, and Miley 1990), and if the outflow disrupts 
enough of the ISM of M82 then its appearance a few hundred Myr in the future 
may be similar to that of NGC 5102 today.

\acknowledgements{It is a pleasure to thank an anonymous referee for suggesting changes 
that greatly improved the paper.}

\clearpage

\begin{table*}
\begin{center}
\begin{tabular}{lc}
\tableline\tableline
Radial Interval & Area Sampled \\
 (kpc) & (arcmin$^2$) \\
\tableline
1 -- 2 & 0.2 \\
2 -- 3 & 1.7 \\
3 -- 4 & 2.5 \\
4 -- 5 & 3.0 \\
5 -- 6 & 3.0 \\
6 -- 7 & 3.0 \\
7 -- 8 & 2.9 \\
8 -- 9 & 2.7 \\
9 -- 10 & 2.7 \\
10 -- 11 & 2.7 \\
\tableline
\end{tabular}
\end{center}
\caption{Area Sampled in Each Annulus}
\end{table*}

\clearpage
\begin{figure}
\figurenum{1}
\epsscale{0.75}
\plotone{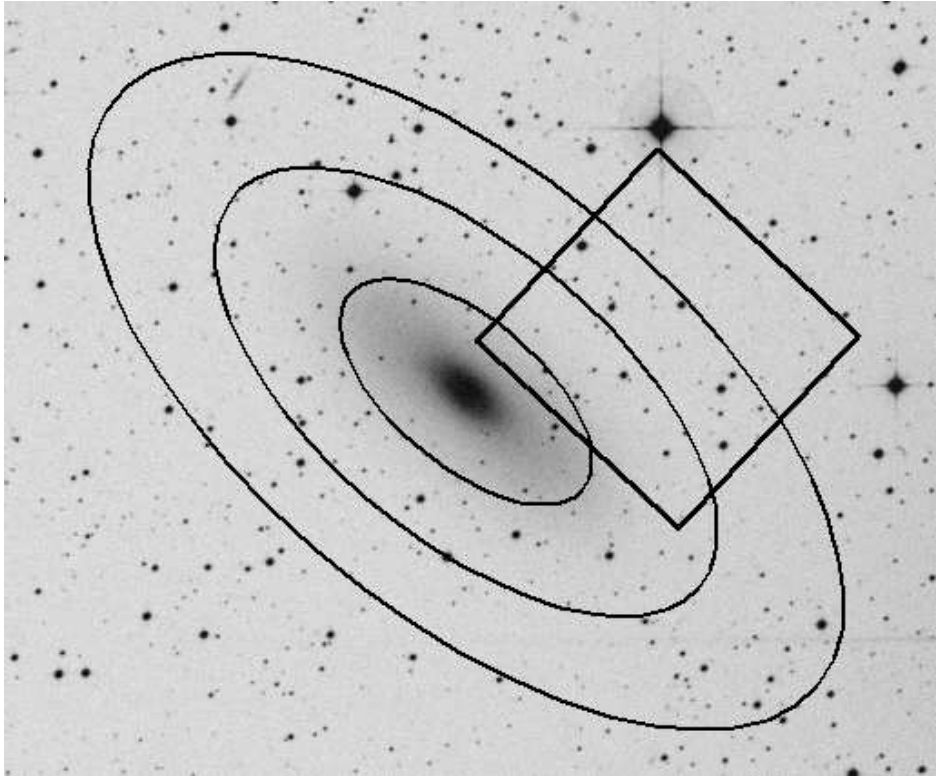}
\caption{The $5.5 \times 5.5$ arcmin$^2$ field that 
was imaged with GMOS is indicated on this section of the red bandpass DSS 
centered on NGC 5102. The ellipses have major axis radii of 
3, 6, and 9 kpc at the distance of NGC 5102. North is at the top, and East is to the left.}
\end{figure}

\clearpage
\begin{figure}
\figurenum{2}
\epsscale{0.75}
\plotone{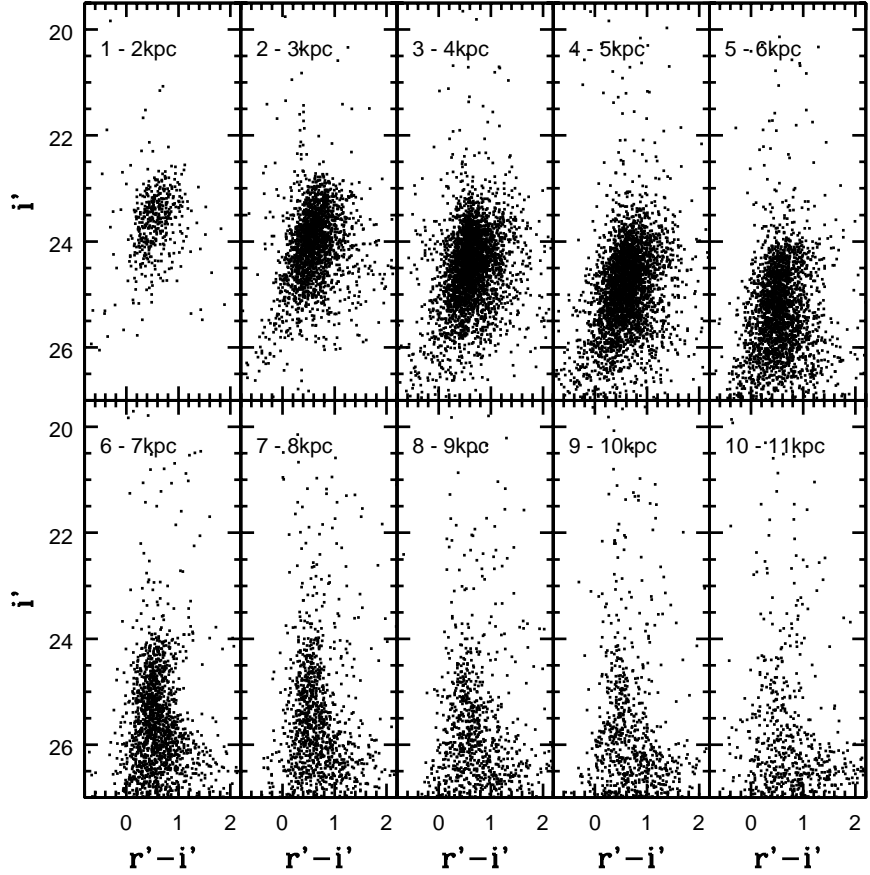}
\caption{The $(i', r'-i')$ CMDs of objects in various radial intervals. The distances 
listed in each panel are along the disk plane, and assume that NGC 5102 is inclined 
to the line of sight by 64.4$^o$ and has a distance modulus of 27.5. Note that a 
coherent red giant branch can be traced out to R$_{GC} \sim 10$ kpc.}
\end{figure}

\clearpage
\begin{figure}
\figurenum{3}
\epsscale{0.75}
\plotone{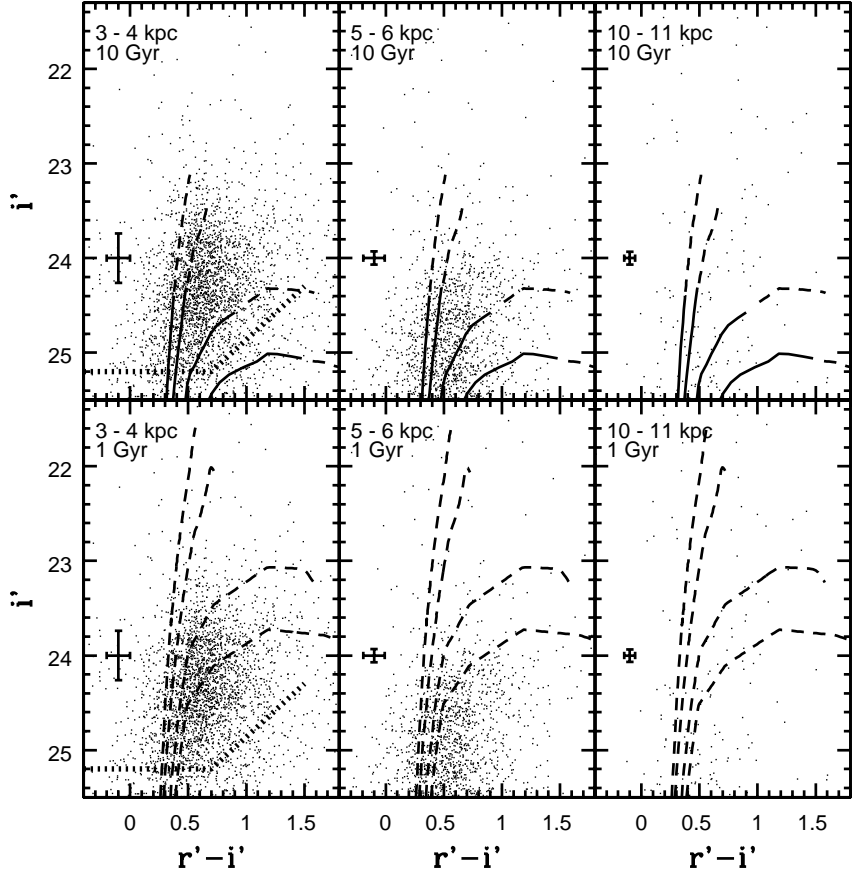}
\caption{Selected isochrones from Girardi et al. (2004) are compared with NGC 5102 
CMDs in this figure. Isochrones with Z = 0.0001, 0.001, 0.004, and 0.008 and 
ages 10 Gyr (top row) and 1 Gyr (bottom row) are shown. The solid lines track 
evolution on the RGB, while the dashed lines track evolution on the AGB. The 
error bars show the random uncertainties predicted by artificial star experiments 
for a star with $i' = 24$ and $r'-i' = 0.5$. The dotted line shows the 50\% 
completeness limit in the 3 -- 4 kpc interval. The 50\% completeness limits 
for the 5 -- 6 kpc and 10 -- 11 kpc intervals fall outside of the plotted magnitude range.}
\end{figure}

\clearpage
\begin{figure}
\figurenum{4}
\epsscale{0.75}
\plotone{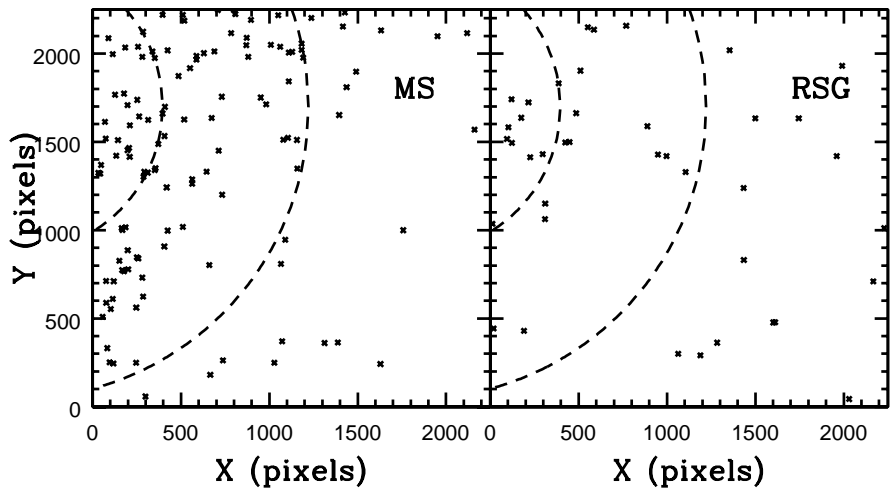}
\caption{The spatial distribution of main sequence stars and RSGs in the western disk 
of NGC 5102. The co-ordinates are in pixel units (0.145 arcsec pixel$^{-1}$), and 
have been de-projected to correct for the inclination of the NGC 5102 disk. 
Main sequence stars have $i' < 25$ and $r'-i' < 0$, whereas 
RSGs have $r'-i'$ between 0.2 and 0.5 and $i'$ between 22.5 and 21.4. The dashed lines 
show circle segments centered on the nucleus of NGC 5102 with radii of 2 arcmin and 
4 arcmin, which are the limits of highest HI concentration in the ring discovered by 
van Woerden et al. (1993).}
\end{figure}

\clearpage
\begin{figure}
\figurenum{5}
\epsscale{0.75}
\plotone{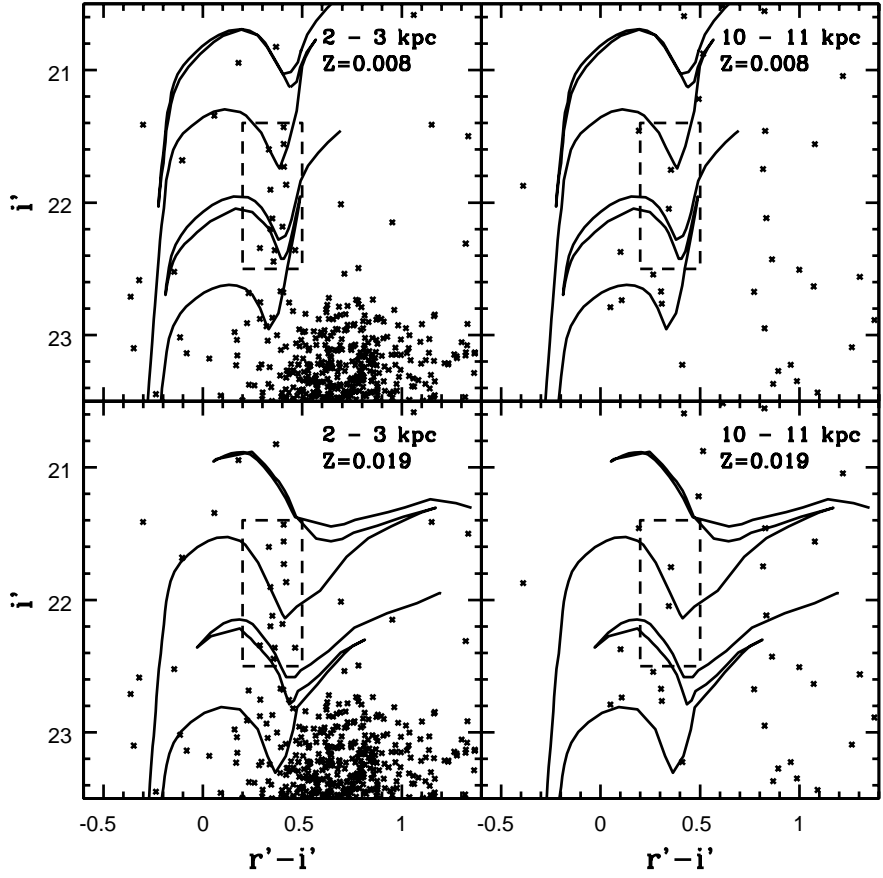}
\caption{Selected isochrones from Girardi et al. (2004) are compared with the CMDs of 
objects with R$_{GC}$ between 2 -- 3 kpc and 10 -- 11 kpc. Isochrones with ages of 20 and 
40 Myr and Z = 0.008 (top row) and Z = 0.019 (bottom row) are shown. The area of the 
CMD that contains the RSG sequence in the 2 -- 3 kpc data is indicated with dashed 
lines. Note that the Z = 0.008 models provide a much better match to the RSG locus 
than the Z = 0.019 models.}
\end{figure}

\clearpage
\begin{figure}
\figurenum{6}
\epsscale{0.75}
\plotone{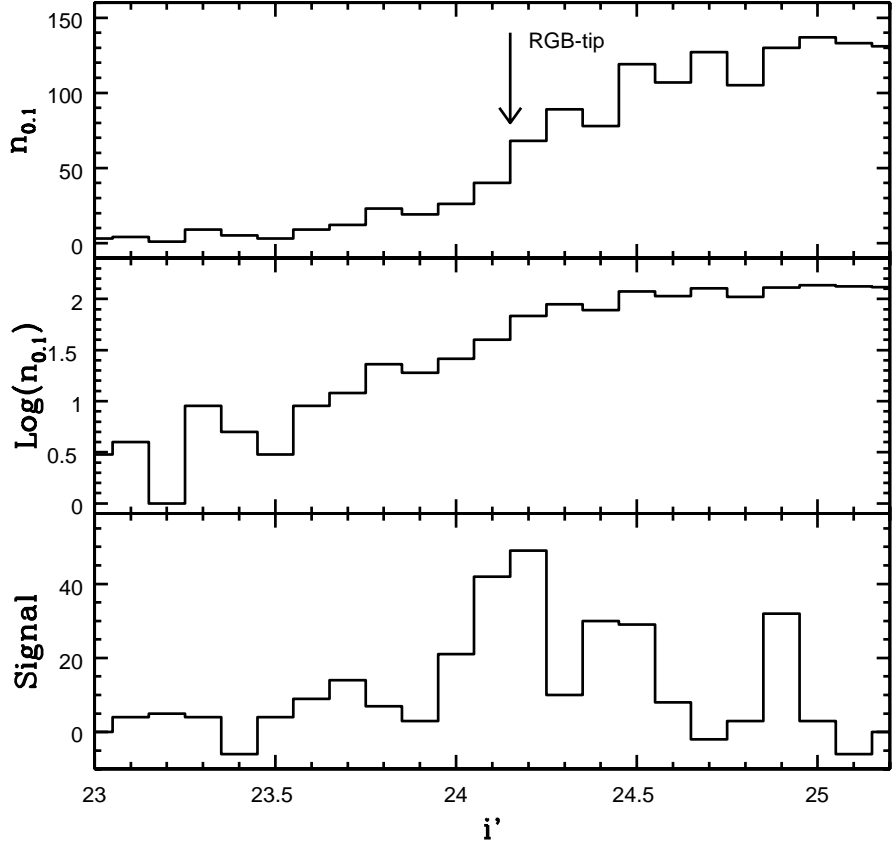}
\caption{The linear and logarithmic $i'$ LFs of stars with $r'-i'$ between 0.3 and 0.6 
are shown in the upper and middle panels of this figure. n$_{0.1}$ is the number of stars 
per 0.1 $i'$ magnitude interval with R$_{GC}$ between 4 and 
7 kpc and $r'-i'$ between 0.3 and 0.6. The result of 
convolving the LF in the upper panel with a three point Sobel edge-detection kernel is 
shown in the lower panel. The filtered signal peaks in the bins centered at $i' = 24.1$ 
and $i' = 24.2$. The weighted mean of the entries in these bins is $i' = 24.15$, which is 
the adopted RGB-tip brightness.}
\end{figure}

\clearpage
\begin{figure}
\figurenum{7}
\epsscale{0.75}
\plotone{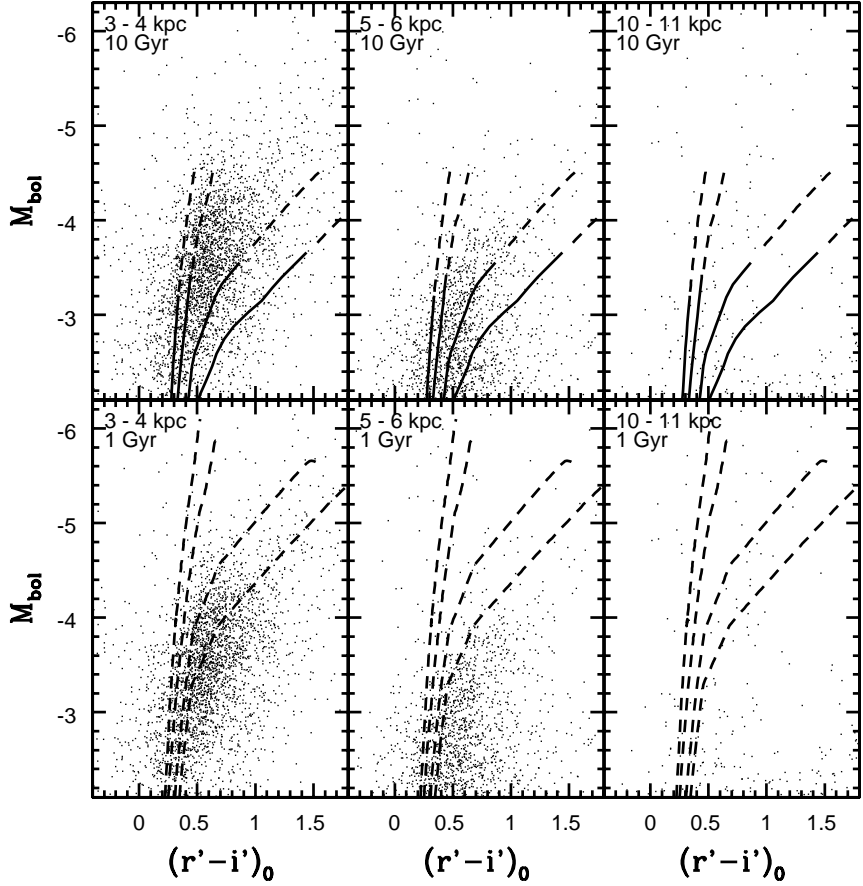}
\caption{The $(M_{bol}, r'-i')$ CMDs of stars in selected radial intervals. M$_{bol}$ 
was computed by assuming (1) $\mu_0 = 27.5$, (2) A$_B = 0.237$, and (3) 
the three-part relation between BC$_{i'}$ and $r'-i'$ that is specified in \S 5.2. 
Isochrones with Z = 0.0001, 0.001, 0.004, and 0.008 
are shown for ages of 10 Gyr (top row) and 1 Gyr (bottom row). The solid lines show 
evolution on the RGB, while the dashed lines show evolution on the AGB.} 
\end{figure}

\clearpage
\begin{figure}
\figurenum{8}
\epsscale{0.75}
\plotone{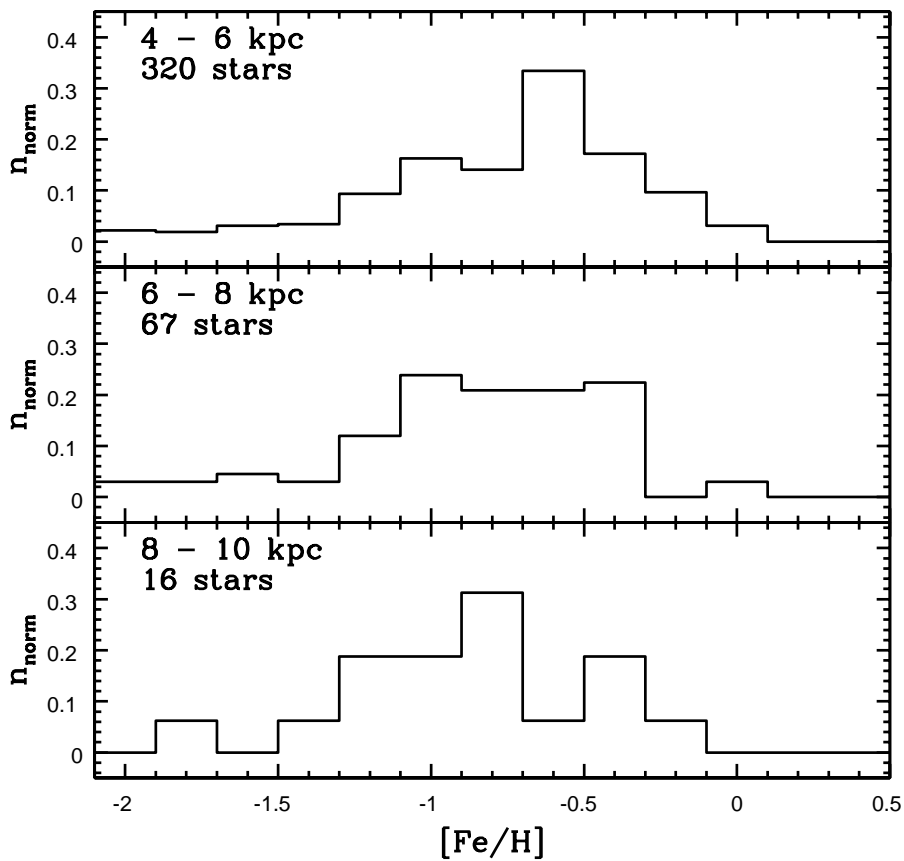}
\caption{The MDFs of stars in NGC 5102 as computed from the $(M_{bol}, r'-i')$ CMDs. 
n$_{norm}$ is the number of stars per 0.2 dex [Fe/H] interval with M$_{bol}$ between 
--2.9 and --3.1, normalized to the total number of stars with [Fe/H] between $-1.3$ 
and $-0.1$ in that radial interval; this latter quantity is listed in each panel. 
The number counts have been corrected for contamination from foreground 
stars and background galaxies by subtracting the MDFs of 
objects with R$_{GC}$ between 10 and 12 kpc.}
\end{figure}

\clearpage
\begin{figure}
\figurenum{9}
\epsscale{0.75}
\plotone{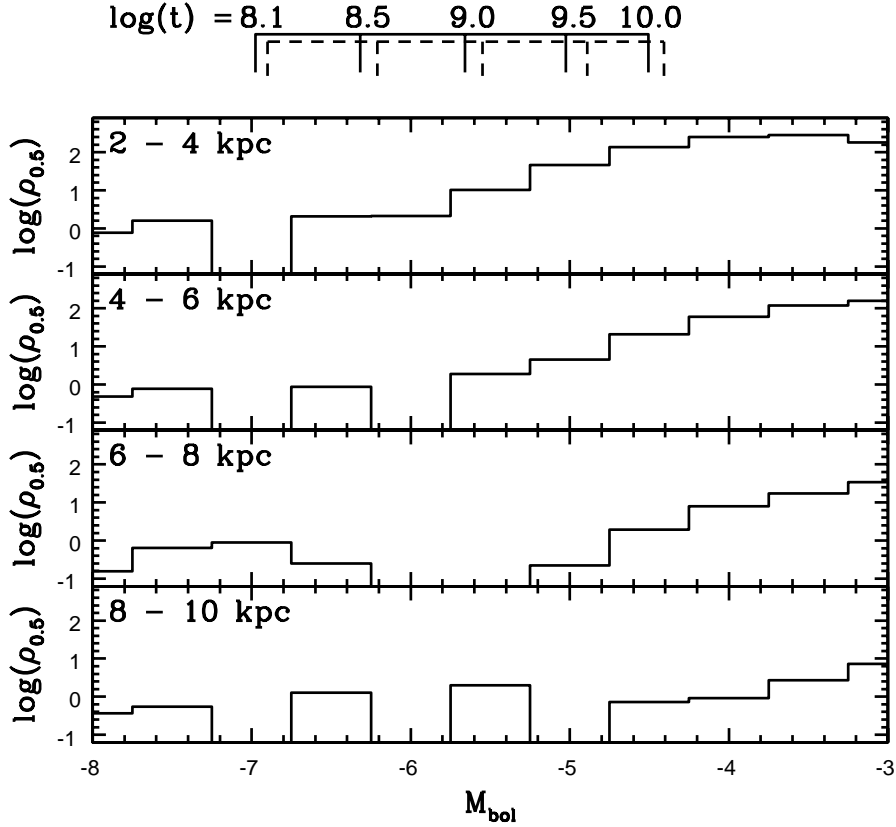}
\caption{The LFs of stars in NGC 5102. $\rho_{0.5}$ is 
the number of stars arcmin$^{-2}$ in each 0.5 M$_{bol}$ interval. Foreground 
stars and background galaxies have been removed statistically by subtracting 
number counts in the 10 -- 12 kpc radial interval. The AGB-tip brightnesses predicted 
from the Z = 0.004 (solid line) and Z = 0.008 (dashed line) models from Girardi 
et al. (2004) for a range of ages are shown at the top of the figure. Note that stars with 
luminosities that are consistent with an age $\leq 1$ Gyr are 
seen throughout much of the NGC 5102 disk.}
\end{figure}

\clearpage
\begin{figure}
\figurenum{10}
\epsscale{0.75}
\plotone{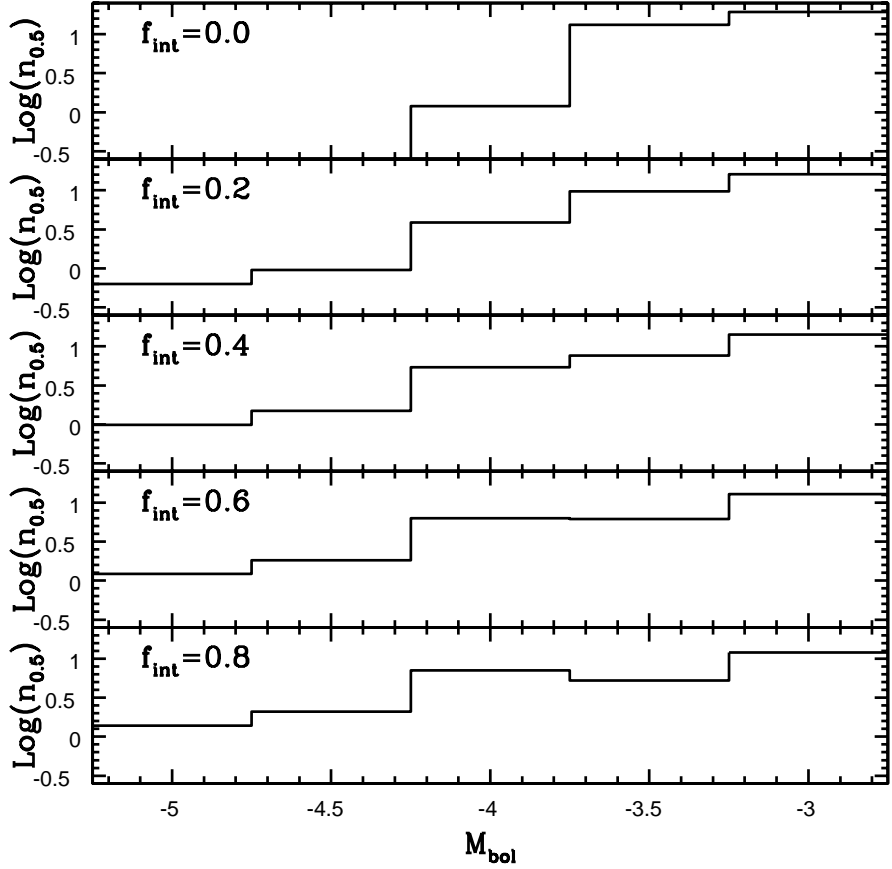}
\caption{A suite of model LFs that were generated by combining the LFs of 
stars in intermediate age LMC clusters and the globular cluster 47 Tuc. f$_{int}$ 
is the fraction of the total stellar mass in intermediate age stars, while 
n$_{0.5}$ is the number of stars in each 0.5 magnitude M$_{bol}$ interval in a system 
with total brightness M$_K = -11$.}
\end{figure}

\clearpage
\begin{figure}
\figurenum{11}
\epsscale{0.75}
\plotone{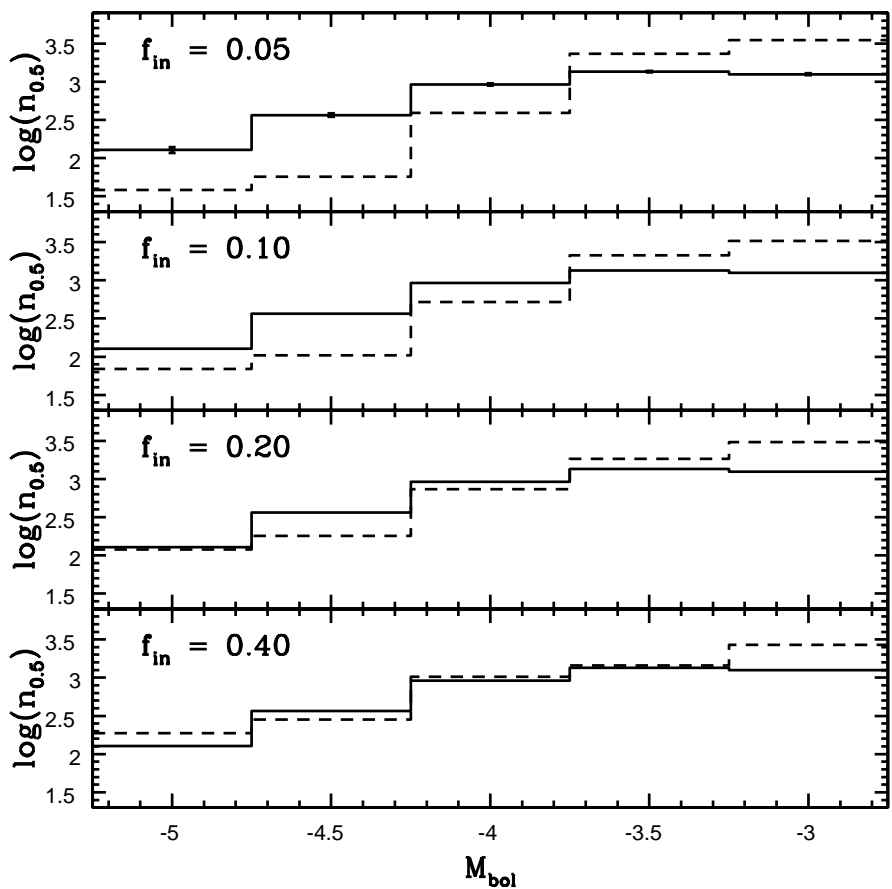}
\caption{The M$_{bol}$ LF of stars with R$_{GC}$ 
between 3 and 5 kpc (solid line) are compared with various model LFs (dashed lines). 
f$_{int}$ is the fraction of the total stellar mass that is made up of 
intermediate age stars. Note that when f$_{int} < 0.2$ the observed LF is shallower 
than the model predictions, whereas when f$_{int} \geq 0.2$ there is reasonable 
agreement between the models and observations.} 
\end{figure}

\end{document}